\documentclass{llncs}

\usepackage{amsmath,url}

\newcommand{\ceil}[1]{\lceil #1 \rceil}
\newcommand{\floor}[1]{\lfloor #1 \rfloor}

\newtheorem{mylemma}{Lemma}
\newtheorem{mytheorem}{Theorem}

\title{New algorithms for binary jumbled pattern matching}

\author{Emanuele Giaquinta\inst{1} \and Szymon Grabowski\inst{2}}

\institute{Department of Computer Science, University of Helsinki, Finland
	  \email{emanuele.giaquinta@cs.helsinki.fi} \and
	  Lodz University of Technology, Institute of Applied Computer Science, Al.\ Politechniki 11, 90--924 {\L}\'od\'z, Poland
	  \email{sgrabow@kis.p.lodz.pl}
}

\begin{document}

\maketitle

\begin{abstract}
  Given a pattern $P$ and a text $T$, both strings over a
  binary alphabet, the \emph{binary jumbled string matching problem}
  consists in telling whether any permutation of $P$ occurs in $T$.
  The indexed version of this problem, i.e., preprocessing a string to
  efficiently answer such permutation queries, is hard and has been
  studied in the last few years. Currently the best bounds
  for this problem are $O(n^2/\log^2 n)$ (with $O(n)$ space and $O(1)$
  query time)~\cite{MR2012} and $O(r^2\log r)$ (with $O(|L|)$ space
  and $O(\log|L|)$ query time)~\cite{BFKL2012}, where $r$ is the
  length of the run-length encoding of $T$ and $|L| = O(n)$ is the
  size of the index. In this paper we present new results for this
  problem. Our first result is an alternative construction of the
  index by Badkobeh et al.~\cite{BFKL2012} that obtains a trade-off
  between the space and the time complexity. It has $O(r^2\log k +
  n/k)$ complexity to build the index, $O(\log k)$ query time, and
  uses $O(n/k + |L|)$ space, where $k$ is a parameter. The second
  result is an $O(n^2 \log^2 w / w)$ algorithm (with $O(n)$ space and
  $O(1)$ query time), based on word-level parallelism where $w$ is the
  word size in bits.
\end{abstract}

\section{Introduction}
\noindent The umbrella term ``approximate string matching'' comprises a plethora of matching models;
one of them, the so-called {\em jumbled string matching}~\cite{DBLP:journals/ipl/ButmanEL04,CFL2009}, does not distinguish between permutations of the pattern string.
This problem, in its decision version, consists in telling if any permutation of a pattern
$P$ occurs in a text $T$, both strings over a finite alphabet.
In the literature, the binary version of this problem, i.e., the one in which the
alphabet for $P$ and $T$ is binary, has been given most attention, and the present article
 is also restricted to this case.
More formally, the {\em binary jumbled string matching} problem can be stated as follows:
We are given a text $T$ of length $n$ over the alphabet $\Sigma = \{0, 1\}$,
the length $m$ of a pattern over the same alphabet, and the number $k$
of symbols 1 in the pattern.
The task is to answer efficiently if there exists a substring of $T$ of length $m$
containing exactly $k$ 1's. The pattern is usually represented as the pair $(m-k,k)$,
called a {\em Parikh vector}.

The online version of this problem can be trivially solved with a $O(n)$ time algorithm
for a single pattern.
It is more interesting, however, to build an index for $T$ making it possible to answer queries
much faster, even in constant time.
The query pattern lengths are arbitrary (can be any values from 1 up to $n$).
Each index-based algorithm can be described by a triple $\langle (f(n), h(n)), g(n) \rangle$,
where $f(n)$ is the preprocessing time, $h(n)$ is the preprocessing space
(which is usually also the size of the resulting index), and $g(n)$ is the query time.
We assume the word-RAM model of computation.

A fundamental observation concerning the binary jumbled string matching is an {\em interval property}:

\begin{mylemma}[\cite{CFL2009}]
If, for a given text $T$ and a pattern length $m$, the
answer is positive for some $(m-k_1, k_1)$ and $(m-k_2, k_2)$, where $k_1 < k_2$,
it is positive also for all $(m-k,k)$ such that $k_1 < k < k_2$.
\end{mylemma}

A practical consequence of this lemma is that it is enough to find the minimum and the maximum
number of 1's for a given pattern length $m$, to be able to give the answers for all $k$ for this $m$.
To avoid a complex notation, we shall identify these values simply as {\em minOne} and {\em maxOne}, respectively.

The first index for binary jumbled string matching, which exploits a
connection with the $(\min, +)$ convolution and whose complexity is
$\langle (O(n^2/\log n), O(n)), O(1) \rangle$, has been independently
discovered by two different groups of
researchers~\cite{DBLP:journals/ipl/MoosaR10,DBLP:conf/fun/BurcsiCFL10}.
Currently the best results for this problem are $\langle (O(n^2/\log^2
n), O(n)), O(1) \rangle$ by Moosa and Rahman~\cite{MR2012} and
$\langle (O(r^2\log r), O(|L|)), O(\log|L|) \rangle$ by Badkobeh et
al.~\cite{BFKL2012}, where $r$ is the length of the run-length
encoding of $T$ and $|L| = O(n)$ is the size of their index structure.
Recently, an interesting result was presented by Cicalese et
al.~\cite{DBLP:conf/cpm/CicaleseLWY12}. They showed how to build an
index for all Parikh vectors of a binary string in $O(n^{1 + \eta})$
time, for any $\eta > 0$, which leaves a chance for false positives,
i.e., may report a Parikh vector not occurring in the string.

In this article we present two novel results.
First, we show that the index from~\cite{BFKL2012} can be easily
modified to become an $\langle (O(r^2\log k + n/k), O(n/k + |L|)),
O(\log k) \rangle$ solution, where $k$ is a trade-off parameter. In
particular, for $k = O(1)$ we obtain an index with $\langle (O(r^2 +
n), O(n)), O(1) \rangle$ complexity.
While the index space is increased, we think that such a trade-off is often
preferable (a more detailed comparison of the results is given in the corresponding section).
Our result improves over the Moosa-Rahman solution if $r = o(n /
\log^{1 + \varepsilon} n)$, for any $\varepsilon > 0$, and over the
solution by Badkobeh et al. if $\sqrt{n/k} = o(r\sqrt{\log r})$ and $k
= r^{O(1)}$.

The second result is an algorithm, based on word-level parallelism,
with $\langle (O(n^2 \log^2 w / w), O(n)), O(1) \rangle$ complexity,
where $w$ is the word size in bits.
It dominates over the $O(n^2 / \log^2 n)$ algorithm
from~\cite{MR2012} if, roughly speaking, $w =
\Omega(\log^{2+\varepsilon} n)$, for any $\varepsilon > 0$. Although
this bound does not hold in practice on current architectures with
$64$-bit word, we believe that it is relevant theoretically. In
particular, occurrences of the so-called wide word assumption date
back to 1995~\cite{DBLP:conf/stoc/AnderssonHNR95}.

Our algorithms, like all the others for this problem, are based on the interval property.
We note that for each interval size it is enough to present a procedure only
for finding {\em maxOne}, since {\em minOne} is equal to {\em maxOne} on negated
input, i.e., where 0's become 1's and vice versa.
Throughout the paper we assume that all logarithms are in base 2.

\section{Basic notions and definitions}
\noindent Let $\Sigma$ denote a finite alphabet
and $\Sigma^m$ the set of all possible
sequences of length $m$ over $\Sigma$. $|S|$ is the length of string
$S$, $S[i], i \geq 0$, denotes its $(i+1)$-th character, and
$S[i\,\ldots\,j]$ its substring between the $(i+1)$-st and the
$(j+1)$-st characters (inclusive). For a binary string $S$ over the
alphabet $\{0,1\}$ we denote with $|S|_0$ and $|S|_1$ the number of
0's and 1's in S, respectively. The Parikh vector of $S$ is the pair
$(|S|_0,|S|_1)$. We say that a Parikh vector $(x,y)$ occurs in a
string $S$ if there exists a substring of $S$ such that its Parikh
vector is equal to $(x,y)$. The run-length encoding of a binary string
$S$ is the sequence $\langle a_1, b_1, a_2, b_2\ldots, a_{\ell},
b_{\ell}\rangle$ of non-negative integers such that
$a_i > 0$ for $i=2,\ldots,\ell$, and $b_i > 0$ for $i=1,\ldots,\ell-1$,
and $S = 0^{a_1}1^{b_1}0^{a_2}1^{b_2}\ldots 0^{a_{\ell}}1^{b_{\ell}}$.
The length $r$ of the run-length encoding of $S$ is then $2\ell - 2\le r\le 2\ell$.
The maximal substrings of a binary string containing only 0's (1's) are
called 0-runs (1-runs).

\section{A variant of the Badkobeh et al. index}

\noindent Let $T$ be a binary string of length $n$ over
$\Sigma=\{0,1\}$ and whose run-length encoding has length $r$. In this
section we present an alternative method to build the Corner Index by
Badkobeh et al.~\cite{BFKL2012} for $T$ that has $\langle (O(r^2\log k
+ n/k), O(n/k + |L|)), O(\log k) \rangle$ complexity, where $k$ is a
parameter. While our construction requires more space, it obtains better
preprocessing and query time. In particular, for $k = O(1)$ it improves the
preprocessing time by a logarithmic factor and yields constant query
time. We briefly recall how the Corner Index works. Let $G(i)$ and
$g(i)$ denote the minimum and maximum number of 1's in a substring of
$T$ containing $i$ 0's, respectively. The following result is a corollary of Lemma $1$:
\begin{mylemma}[cf.~\cite{BFKL2012}]
  Given a Parikh vector $(x,y)$ and a binary string $T$ with
  associated functions $G$ and $g$, $(x,y)$ occurs in $T$ iff $G(x)\le
  y\le g(x)$.
\end{mylemma}
Hence, to be able to know whether any Parikh vector occurs in $T$ it
suffices to compute the functions $G$ and $g$. Since $G$ and $g$ are
monotonically increasing, they can be encoded by storing only the
points where they increase. Let $L_G$ and $L_g$ be the sets of such
points for $G$ and $g$, respectively. The Corner Index of $T$ is the
pair $(L_G, L_g)$. In accordance with the notation from~\cite{BFKL2012},
let us use the symbol $L$ denoting the whole Corner Index, and
obviously we have $|L_G| + |L_g| = |L|$. From now on we consider the
construction of $L_G$ only, since the case of $L_g$ is analogous and
has the same space and time bounds. The set $L_G$ is defined as
$$
L_G = \{(i,G(i))\ |\ G(i) < G(i+1)\}\,.
$$
The function $G$ can be reconstructed from $L_G$ based on the relation
$G(x) = G(r(x))$, where
$$
r(x) = \min \{i\ |\ i\ge x\wedge (i,G(i))\in L_G\}\,.
$$
Let $\Pi^{rle}(T)$ be the set of Parikh vectors of all the substrings
of $T$ beginning and ending with full 0-runs. The authors
of~\cite{BFKL2012} showed that, in order to build $L_G$, it is enough
to compute $\Pi^{rle}(T)$, as $L_G\subseteq \Pi^{rle}(T)$.
Formally, the set $L_G$ corresponds to the set of maximal elements of
the partially ordered set $(\Pi^{rle}(T),\triangleright)$, where the
relation $\triangleright$ is defined as
$$
(x,y)\triangleright (x',y')\iff (x,y)\neq (x',y')\wedge x\ge x'\wedge y\le y'\,.
$$
If $(x,y)\triangleright (x',y')$, $(x,y)$ is said to dominate $(x',y')$.
The set $\Pi^{rle}(T)$ can be computed efficiently on the run-length
encoding of $T$ in time $O(r^2)$. The total time complexity of the
procedure to build the Corner Index is $O(r^2\log r)$, since for each
such Parikh vector the algorithm performs one lookup and at most one
insertion and deletion in a balanced tree data structure whose size is
at most $r^2$. We now show how to achieve a trade-off between the
space and time complexity. We divide the interval $[1,|T|_0]$ into
sub-intervals of $k$ elements, such that length $i$ is mapped to the
$\floor{i/k}$-th interval, for $i=1,\ldots,|T|_0$. For each interval
$I_j=[(j-1)\cdot k+1, j\cdot k]$, for $j=1,\ldots,\ceil{|T|_0/k}$, we
maintain a balanced binary search tree (BST) in which we store the set of
maximal elements of $(\Pi^{rle}(T),\triangleright)$ with first component
in $I_j$. If this set is empty, we store
the element $((j-1)\cdot k+1, y)$, where $y$ is the second component of the
element in the previous interval with largest first component.
The value of $G$ on a given point can then be computed in time $O(\log k)$
by doing a lookup in the tree of the corresponding interval. The size
of this index is $O(n/k + |L|)$, as we add at most one redundant pair
per (empty) interval.

We now describe how this index can be built in time $O(r^2\log k + n / k)$.
The construction is divided into two similar steps. We use an array
$V$ of $\ceil{n/k}$ pointers, where each slot points to a BST.
During the first step, we insert each element $(x,y)$ from
$\Pi^{rle}(T)$ into the BST pointed by $V[\floor{y/k}]$ if it is not
dominated by any existing pair and, after an insertion, we also remove
all the existing elements dominated by it.
This procedure is analogous to the one used in the original Corner Index to add an element to the index.
In this phase we
conceptually divide the interval $[1,|T|_1]$ into sub-intervals
($1$-intervals) and partition the pairs according to their second
component. At the end of this step, in each BST we have a superset of
the set of maximal elements of $(\Pi^{rle}(T),\triangleright)$ of the
corresponding $1$-interval. In particular, all the pairs that are
dominated by a pair that belongs to a different $1$-interval are not
removed. The second step of the preprocessing removes these elements
and builds the final index. To this end, we use another array $V'$
defined as $V$. We iterate over $V$ from left to right maintaining in
an integer $x_{\max}$ the maximum first component of any pair
belonging to the BSTs already processed. For each $j=1,\ldots,\ceil{|T|_1/k}$,
we insert all the elements $(x,y)$ of the BST $V[j]$ such that $x >
x_{\max}$ into the BST $V'[\floor{x/k}]$.

It remains to prove that all the
dominated pairs are removed using this procedure. Let $(x,y)$ and
$(x',y')$ be two elements of $\Pi^{rle}(T)$ such that
$(x,y)\triangleright (x',y')$. We distinguish two cases: if the two
elements map onto the same $1$-interval, then only $(x,y)$ remains at
the end of the first pass; otherwise we have $y < y'$ and so we
process $(x,y)$ before $(x',y')$ during the second pass. Hence, when
we process $(x',y')$ it must hold that $x_{\max}\ge x$, so $(x',y')$
is skipped.

Clearly, by definition of the $\triangleright$ relation, the size of
any tree is bounded by $k$ in both phases. Hence, the total procedure
can be performed in time $O(r^2\log k + n/k)$ on the run length
encoding of $T$. To handle the case of trees that are empty, it
suffices to keep track, during the second pass, of the element with
largest first component for each non-empty interval. Then, we perform a
subsequent pass over the intervals, in time $O(n/k)$, and each time we
visit an empty interval $I_j$ we add to it the element $((j-1)\cdot k
+ 1, y)$, where $y$ is the second component of the largest element in
the last non-empty interval visited. We thus obtain the following theorem:

\begin{mytheorem}
  Given a binary string of length $n$ and whose run-length encoding
  has length $r$, we can build an index for jumbled pattern matching
  in time $O(r^2\log k + n/k)$ using $O(n/k + |L|)$ space which
  answers queries in time $O(\log k)$, for any parameter $1\le k\le n$, where
  $|L| = O(n)$.
\end{mytheorem}

Observe that, for $k = O(1)$, we obtain an index with $\langle (O(r^2 + n),
O(n)), O(1)\rangle$ complexity, as promised at the beginning of this
section; instead, for $k=\log n$, we get $\langle (O(r^2\log\log n +
n/\log n), O(n/\log n + |L|)), O(\log\log n)\rangle$, i.e., a smaller
index with sublogarithmic query time. Note that for $k=n$ we obtain
the original Corner Index.

Concerning the index build time, our result dominates over the
Moosa-Rahman solution if $r = o(n / \log^{1 + \varepsilon} n)$, for any
$\varepsilon > 0$, and over the solution by Badkobeh et al. if
$\sqrt{n/k} = o(r\sqrt{\log r})$ and $k = r^{O(1)}$.

It is also possible to replace the balanced binary search trees with a
more advanced data structure to achieve even better query time. In
particular, the exponential search
tree~\cite{DBLP:journals/jacm/AnderssonT07} is a search data structure
that requires $O(\log\log n \frac{\log\log U}{\log\log\log U})$ time
for all the operations, where $U$ is an upper bound on the largest
key. If, for each interval $I_j$, we represent a pair $(x,y)$ in the
interval using the key $x - ((j-1)\cdot k + 1)$, then it holds that
the largest key is $k$ and we thus obtain an index with $\langle
(O(r^2\frac{(\log\log k)^2}{\log\log\log k} + n / k), O(n/k + |L|)),
O(\frac{(\log\log k)^2}{\log\log\log k})\rangle$ complexity.

\section{An algorithm based on word-level parallelism}

\noindent
In this section we present an algorithm based on word-level
parallelism. Each interval size is processed separately, and the
(packed) text is scanned from left to right in chunks of $w$ bits,
where $w$ is the machine word size. The text is first packed in $O(n
\log w / w)$ words so that each symbol ($0$ or $1$) is represented
with $f = 1 + \log w$
bits. We will refer to such sequences of bits encoding a symbol as
{\em fields}.
The packed text can be easily computed by setting, for each position
$i$ in $T$, the value of the field with index $(i\mod k)$ in the
$\floor{i / k}$-th chunk to $T[i]$.
Let $k = \lfloor w / f\rfloor$ be the number of symbols
in a chunk. We denote with $C[i]$ the $i$-th symbol of chunk $C$.
Given a
length $1\le l\le n$, the idea is to slide a window of length $l$ over
the packed text, i.e., spanning $\lceil l / k\rceil$ chunks, and
compute the number of 1's in each alignment.
We assume that $l\ge k$ since, if $l < k$, we can
compute {\em maxOne} for $l$ by working with chunks of $l$ symbols rather
than $k$ (with a negligible overhead).
Note that if $l$ does not divide $k$, the
window spans only the prefix of length $l\bmod k$ of the last chunk,
which can be obtained by masking the rightmost $k - l\bmod k$ fields with a bitwise \texttt{and}. For simplicity,
we assume that $l$ divides $k$. To begin with, we compute the maximum
number of 1's in the first alignment, i.e., in the first $l / k$
chunks of the text. We then slide the window over the packed text by
extending the window by one chunk to the right and reducing it by one
chunk from the left, i.e., for a window beginning at position $i$ we
perform a transition from $T[i \ldots i+l-1]$ to $T[i + k \ldots i
+ k + l - 1]$. At each iteration of this process, we compute the
maximum number of 1's among the $k$ new shifts of the window in time
$O(\log w)$ and update {\em maxOne} if needed. When the current window
is moved to the right by one chunk (i.e., by $k$ symbols), two text
chunks are affected, the one including the symbols that fall off the
window and the one with the symbols that enter it. Let us denote these
chunks with $C_1$ and $C_2$, respectively. It is not hard to see that
the maximum number of 1's among the new shifts is equal to $ones +
\max_h\{\sum_{i=0}^{h-1} C_2[i] - \sum_{i=0}^{h-1}C_1[i]\}$, for $h=1,\ldots,k$, where
$ones$ is the number of 1's in the chunks spanned by the previous
alignment. Hence, to update {\em maxOne}, we
should know if there is at least one prefix of length $h$ such that
the difference between the sums of $C_2[0 \ldots h-1]$ and $C_1[0
\ldots h-1]$ is positive and, if the answer is affirmative, what the
maximum difference of sums over such equal-length prefixes of $C_2$
and $C_1$ is. Note that we do not have to know the prefix length that
maximizes this difference. We now show how to compute this value for
two chunks of $k$ symbols in time $O(\log w)$ by exploiting the relation with the
problem of computing the prefix sums of a sequence of size $k$ of $f$-bit
numbers.
We proceed as follows: first, we compute, in constant time, a new word $C'$ such that
$$
C'[i] = C_2[i] + 1 - C_1[i]\,,
$$
for $i=0,\ldots, k-1$, by first adding to $C_2$ a word where all the fields have value $1$ and then subtracting $C_1$.
The word $C'$ holds the differences between the fields in $C_1$ and
$C_2$ with equal position augmented by one unit. Observe that we
defined $C'$ in this way so as to not let the fields obtain negative
values, i.e., all the values in $C'$ are non-negative ($0$, $1$, or $2$).
To this end, the addition must also be performed before the subtraction.
We avoid negative numbers because their encoding is dependent on the
model used by the machine. In particular, in the most common two-complement
arithmetic, a negative value in a field would result in changes to
other fields, thereby making the result incorrect.

Then, we compute the prefix sums of the $k$ symbols of $C'$ in time
$O(\log w)$ by adapting the algorithm by Hillis and Steele~\cite{HS1986} to
compute the prefix sums of an array in parallel. The adaption is
straightfoward: we perform $\ceil{\log k}$ passes over $C'$ where the $i$-th
pass computes the value
$$
C''_i = \begin{cases}
  C''_{i-1} + (C''_{i-1}\ll (2^i\times f)) & \text{if } i > 0 \\
  C' & \text{otherwise} \\
\end{cases}
$$
Note that $\ll$ is a logical shift.
Observe that this step can be computed with a constant number of words
of space. Moreover, no overflow can occur since the largest sum has
value $2k$, while we have $1 + \log w$ bits to store each number. In this way we almost obtain in the word $C''$ the prefix
sums of the differences. More precisely, we have
$$
C''[h] = \sum_{i=0}^{h} (C_2[i] - C_1[i]) + h + 1\,,
$$
for $h=0,\ldots,k-1$, i.e., the $h$-th difference is off by a value of $h+1$ with respect to
the actual value. To fix them, we observe that any value $C''[h]$ that
is smaller than $h+1$ is not interesting, as the real difference is
negative in this case. In what follows, care must be taken not to let
the fields obtain negative values. We generate a word of increments
$I$ with the following field values: $1, 2, \ldots, k$. We find in
parallel the maxima of the pairs of corresponding fields from $C''$
and $I$, using the technique from~\cite{PS1980} that works in constant
time.
The idea is to compute a word in which the top bit of each field $h$
is $1$ if $\max(C''[h], I[h]) = C''[h]$ and $0$ otherwise, for
$h=0,\ldots,k-1$, and then use it to extract the correct fields from
both words.
This is why the fields have exactly $1 + \log w$ bits. As a result,
the new value of $C''$ has some fields taken from the ``old'' $C''$,
and some from $I$. Finally we subtract $I$ from $C''$; clearly, no
field will obtain a negative value, and zeros will appear in the
fields corresponding to prefix sum differences less than or equal to
zero.

The final phase is to find quickly the maximum field value in the
whole word. Note that we can spend $O(\log w)$ time for this step
without changing the overall time complexity. We can obtain the
desired time complexity with the following algorithm: we logically
divide $C''$ in two halves and initialize two words with the first and
last $k/2$ fields of $C''$, respectively. Then, we compute in parallel
the maxima between the two words in constant time, using again the
procedure from~\cite{PS1980}. Note that the resulting word logically
contains $k/2$ fields only. We recursively repeat this process up to
$\ceil{\log k} = O(\log w)$ passes. The last word will contain exactly
one field, the maximum difference.

We spend $O(\log w)$ time to process $k$ symbols of the input data,
hence the speed-up over the naive algorithm is by factor $
\Theta(w/\log^2 w)$, which gives overall $O(n^2 \log^2 w / w)$ time
complexity. We thus obtain the following theorem:

\begin{mytheorem}
  Given a binary string of length $n$, we can build an index for
  jumbled pattern matching in time $O(n^2\log^2 w / w)$, in the
  word-RAM model, using $O(n)$ space which answers queries in
  constant time.
\end{mytheorem}

This result dominates over the $O(n^2 / \log^2 n)$ algorithm
from~\cite{MR2012} if, roughly speaking, $w =
\Omega(\log^{2+\varepsilon} n)$, for any $\varepsilon > 0$.

\section{Acknowledgments}

We thank the anonymous reviewers for helpful comments.

\bibliographystyle{abbrv}
\bibliography{bjpm}

\end{document}